\documentclass[aps]{revtex4}
\usepackage{amssymb,amsmath,bm,enumerate,graphicx,color,subfigure}
\begin{document}
\title{Trajectory statistics of confined L\'{e}vy flights and Boltzmann-type equilibria}
\author{Mariusz  \.{Z}aba,  Piotr Garbaczewski  and  Vladimir Stephanovich}
\affiliation{Institute of Physics, University of Opole, 45-052 Opole, Poland}
\begin{abstract}
We analyze  a specific class of   random  systems   that  are  driven  by  a symmetric  L\'{e}vy stable noise, where Langevin representation is absent. In view of   the  L\'{e}vy noise sensitivity to environmental inhomogeneities,
the pertinent  random motion asymptotically sets down at  the  Boltzmann-type equilibrium, represented by a  probability density function (pdf)  $\rho _*(x) \sim \exp [-\Phi (x)]$. Here, we infer pdf $\rho (x,t)$ based on
 numerical path-wise simulation of the underlying jump-type process. A priori  given data are jump transition rates entering the master equation  for  $\rho (x,t)$  and its target  pdf  $\rho _*(x)$.
  To simulate the above processes, we construct a suitable modification of the Gillespie algorithm,  originally invented  in  the  chemical kinetics context. We exemplified our algorithm simulating different jump-type processes
  and discuss the dynamics of real physical systems where it can be useful.
\end{abstract}
\maketitle

\section{Introduction}

Despite many attempts to pin down the essential features of dynamics and relaxation in random systems, the problem is still far from its complete solution. It turns out that  L\'{e}vy flight models are
adequate for the  description of different random systems ranging from the motion of defects in disordered solids to the dynamics of assets in stock markets, see, e.g. \cite{vk}.
They are especially useful to model the random systems on the semi-phenomenological, mesoscopic level, when the (often unknown) details of their microscopic random behavior are substituted by  a  properly tailored
(e.g. based on experimental data) (Gaussian or L\'{e}vy) noise. Paradoxically, in disordered solids, the noise can promote order and organization, switching them between different equilibrium states.  The latter
 situation emerges in disordered ferroelectrics, where the fluctuations of order parameter (spontaneous polarization) give rise to self-localization of charge carrier, generating a fluctuon,   an
 analog of well-known polaron in disordered substance \cite{stef97}. These fluctuons make a substantial contribution into conductivity and optical properties of disordered dielectrics.

Many  random processes admit  a description based on stochastic differential equations.
In such case there is a routine  passage  procedure  from microscopic   random variables to macroscopic (statistical  ensemble) data.
 The latter are   encoded in the   time evolution of an associated probability density function (pdf)  which  is  a solution of  a deterministic  transport equation.
A paradigmatic example  is  the  so-called Langevin modeling of diffusion-type and jump-type processes.
The presumed microscopic model of the dynamics in external force fields  is provided by the Langevin (stochastic)  equation whose
direct  consequence is   the   Fokker-Planck   equation, \cite{risken} and \cite{fogedby}.
 We note that in case of jump-type processes the familiar  Laplacian (Wiener noise generator) needs
  to be replaced by a suitable pseudo-differential operator (fractional Laplacian, in case of a symmetric   L\'{e}vy-stable noise).

 We pay a particular attention to jump-type processes which are omnipresent  in Nature (see \cite{metzklaf} and references therein).  Their characterization is  primarily provided by jump  transition
 rates between different states of the system under consideration.   In the present paper,   our major  focus is  on a  specific  class of random systems which are   plainly incompatible with
 a straightforward Langevin modeling of jump-type processes and,  as such,  are seldom  addressed in the literature.

To this end we depart  from the  concept, coined in an  isolated  publication  \cite{deem},  of
L\'{e}vy flights-driven models of disorder that, while at equilibrium, do  obey detailed balance.
The corresponding  research  line  has been effectively  initiated in Refs.  \cite{brockmann}-\cite{belik}.
It has  next  been expanded in various directions,   with
a special  emphasis  on so-called L\'{e}vy-Schr\"{o}dinger  semigroup reformulation of  the original    probability density function
(pdf)  dynamics,  \cite{vilela1,vilela2,geisel}  and
 \cite{olk}-\cite{gar},  c.f. also \cite{lorinczi,vilela1,vilela2}.  We note in passing that the familiar
  Fokker-Planck  equation can be equally well   reformulated in terms of the  Schr\"{o}dinger semigroup and this property is universally valid in
 the  standard theory of Brownian motion,  \cite{risken,lorinczi}. Its generalization to L\'{e}vy   flights is neither  immediate nor obvious.
  It is   often considered in the  prohibitive vein following \cite{klafter,cohen}.

In fact, in relation to  L\'{e}vy flights,  a  novel   fractional  generalization  of the   Fokker-Planck equation has been introduced in Refs. \cite{brockmann}-\cite{belik}
    to handle systems that are randomized  by symmetric  L\'{e}vy-stable  drivers. In this case,  contrary to  the  popular lore  about properties of
     (Langevin-based) L\'{e}vy processes (c.f. Refs. \cite{klafter}-\cite{dubkov} and \cite{cohen}),  the pertinent   random systems  are allowed to
      relax to   (thermal) equilibrium states of a standard Boltzmann-Gibbs form.

      The  underlying  jump-type processes, in the stationary (equilibrium)  regime,
 respect the  principle of  detailed balance  by construction \cite{gar,gar3}.    Their distinctive feature,  if compared with  the standard Langevin modeling
 of    L\'{e}vy flights,
is that  they have a built-in response  {\it not}  to external forces {\it  but rather to} external force potentials.
These potentials are interpreted  to form  confining "potential landscapes" that are specific to the environment.  L\'{e}vy  jump-type
 processes  appear to be particularly sensitive to environmental inhomogeneities, \cite{brockmann,gar2}.

  We note in passing that confining L\'{e}vy flights   resolve themselves to the problem of truncated distributions, which can be addressed in many ways.
   beside a simple cut-off,   by   considering   fast falling tails.  The random walk  theory  predominantly  employs to the master equation as the major  tool
   to quantify random  motion.   Inhomogeneous  transition rates have been used in the  literature before and the  emergence  of non-trivial target distributions
    has been demonstrated, c.f. \cite{srokowski}.

L\'{e}vy  flights  are pure jump (jump-type) processes. Therefore, it seems useful  to indicate  that various model realizations of  standard
jump processes (jump size is bounded from below and above) can be thermalized  by means of a specific scenario of an energy exchange with the thermostat. It is
based on the {\it  principle of detailed balance}.
We have discussed this issue in some detail before \cite{gar} along with an extension of  this conceptual framework  to L\'{e}vy-stable processes.
Not to reproduce   easily   available   arguments of past publications, we shall be very rudimentary in our motivations, see however \cite{gar3}.

We quantify a  probability density evolution, compatible with a  jump-type process on $R$ (this limitation may in principle be lifted in favor of $R^n$),
 in terms of the master  equation:
\begin{equation}
\partial_t\rho(x,t)=\int\limits_{\varepsilon_1\leq |x-y|\leq \varepsilon_2} [w_\phi(x|y)\rho(y,t)-w_\phi(y|x)\rho(x,t)]dy,\label{l1}
\end{equation}
where $\varepsilon_1$ and $\varepsilon_2$ are, respectively, the lower and upper bounds of jump size and
\begin{eqnarray}
w_\phi(x|y)&=&C_\mu\frac{\exp[(\Phi(y)-\Phi(x))/2]}{|x-y|^{1+\mu}},\nonumber \\ \nonumber \\
C_\mu&=&\frac{\Gamma(1+\mu)\sin(\pi\mu/2)}{\pi} \label{l2}
\end{eqnarray}
is the  jump   transition rate  from  $y$  to  $x$.  We stress that  $w_\phi(x|y) $ is a  non-symmetric function of $x$ and $y$.

An implicit Boltzmann-type weighting involves a square root of a target pdf  $\rho _*(x) \sim  \exp[-\Phi  (x)]$ and  accounts for
the  a priori  prescribed   "potential landscape" $\Phi  (x)$  whose confining features affect the  jump-type process.
What  matters is  a relative  impact   of a confinement strength   of $\Phi (x)$  (level of attraction, see Ref. \cite{belik})
 upon jumps of  the size  $|x-y|$,    both at  the   point  of origin  $y$ and that of  destination   $x$.
  In principle, $\Phi(x)$ may be an arbitrary function that secures a  $L^1(R)$  normalization of $\exp(-\Phi(x))$.
  In this case, the resultant pdf $\rho _*$  is a  stationary solution of the transport equation
   (\ref{l1}) with  unbounded jump length, e.g.  $\varepsilon_1 \to 0$   and  $\varepsilon_2 \to\infty$.

 We note that the presence of lower and upper bounds of the jump size $\varepsilon_{1,2}$, that are necessary for an implementation of  numerical algorithms,
 enforce a truncation of  the jump-type  process  (without any cutoffs) to a standard  jump process. The transition rates of the latter,  however, are ruled by
 L\'{e}vy measures of symmetric  L\'{e}vy stable noises with  $\mu \in (0,2)$.
 A lower bound for  the jump size  is usually removed while evaluating the corresponding integrals in the  sense of their Cauchy principal values.
An upper bound is less innocent and its effects need to be controlled by long tailed  pdfs which  stands for a distinctive
feature of L\'{e}vy flights, see a discussion of L\'{e}vy stable limits of step  processes in Ref.  \cite{olk}.
There is also pertinent discussion of a long time behavior of   (unconfined, e.g. free)  truncated L\'{e}vy flights in Ref. \cite{mantegna}.

In contrast to procedures based on the Langevin modeling of L\'{e}vy flights in external force fields, \cite{fogedby,chechkin,dubkov},
 there is no known path-wise approach underlying  the transport equation  (\ref{l1}). With  no  direct  access to  sample  trajectories
  of the stochastic process in question,  a  method must be devised to generate random paths directly
   from jump transition rates (\ref{l2}).  The additional requirement here is that we set
    a priori a  "potential landscape" $\Phi (x)$ for a  chosen  jump-type (symmetric L\'{e}vy stable) noise driver.

The outline of the paper is as follows.   First we describe our modification of the Gillespie algorithm which entails  a numerical generation
of random paths for the dynamics determined by Eqs  (\ref{l1}) and  (\ref{l2}).  Next the statistics of random paths is addressed and
various accumulated data are analyzed with a focus on  inherent compatibility issues.

We  analyze  generic  (Cauchy, quadratic Cauchy)  and non-generic  (Gaussian  and locally periodic)  examples of target pdfs  for the jumping dynamics.
 Random paths are generated in conjunction with  representative L\'{e}vy stable drivers, like e.g. those  indexed by  $\mu =1/2, 1,  3/2$.
 Their qualitative typicality is  emphasized.

Statistical data, acquired from our modification of Gillespie algorithm, have been employed  to generate the dynamical patterns
of  behavior $\rho (x,t) \to \rho _*(x)$.   Effectively,  that entails a  path-wise   reconstruction  of  the solution  $\rho (x,t)$
 of the master equation (\ref{l1}),  (\ref{l2}), or - in case  this solution it is available prior to  trajectory  simulations -
 to   verify a  compatibility of the transport (master)  equation   (\ref{l1}),  (\ref{l2})  and  its  underlying   path-wise representation.
 All that comes from  the   predefined  knowledge of the target pdf and non-symmetric (biased)  jump transition rates.

\section{Modified Gillespie algorithm.}

Here we adopt   \cite{sokolov} (and  properly adjust to handle L\'{e}vy  flights) basic tenets of  so-called
Gillespie's algorithm \cite{gillespie,gillespie1}.
Originally, this algorithm had been devised to simulate  random properties of coupled chemical reactions.
The advantage of the algorithm is that it permits to generate  random trajectories of the  corresponding stochastic process directly from its (jump) transition rates,
 with  no need for any  stochastic differential equation and/or its explicit solution.  We emphasize that this  feature of Gillespie's algorithm is vitally important, since
  Langevin modeling is not  operational in our framework.

We   rewrite Eq. (\ref{l1})  in  the form ($x-y=z$)
\begin{equation}
\partial_t\rho(x,t)=\int\limits_{\varepsilon_1\leq|z|\leq \varepsilon_2} \biggl[w_\phi(x|z+x)\rho(z+x,t)- 
-w_\phi(z+x|x)\rho(x,t)\biggr]dz.\label{l3}
\end{equation}
To construct a reliable path generating algorithm consistent with Eq. (\ref{l3}) we first note that
chemical reaction channels in the original Gillespie's  algorithm  may be re-interpreted as jumps  from one spatial point
 to another, like transition channels in the spatial  jump process. An obvious provision is that the set of possible chemical reaction channels is discrete,
 while we are interested in continuous set of all admissible jumps from a chosen point of origin  $x_0$. With a  genuine computer simulation in mind,
 we must respect standard numerical assistance  limitations. Surely we cannot  admit all conceivable jump sizes. As well, the number of destination points,
  even if potentially enormous, must remain finite for any fixed point of origin.

Our modified version of the  Gillespie's algorithm, appropriate for handling of spatial  jumps is as follows \cite{available}:
\begin{enumerate}
\item Set time  $t=0$ and the point of origin  $x=x_0$.
\item Create the set of all admissible  jumps from $x_0$ to $x_0+z$ that is compatible with the transition rate  $w_\phi(z+x_0|x_0)$. \\
\item Evaluate
\begin{eqnarray}
&&W_1(x_0)=\int_{-\varepsilon_2}^{-\varepsilon_1}w_\phi(z+x_0|x_0)dz, \nonumber \\
&&W_2(x_0)=\int_{\varepsilon_1}^{\varepsilon_2}w_\phi(z+x_0|x_0)dz \label{l4}
\end{eqnarray}
and  $W(x_0)=W_1(x_0)+W_2(x_0)$.
\item Using  a random number generator draw  $p\in[0,1]$  from a uniform distribution.
\item Using above $p$   and identities
\begin{equation}
\left\{
  \begin{array}{ll}
    \int\limits_{-\varepsilon_2}^{b}w_\phi(z+x_0|x_0)dz=p W(x_0), & \hbox{$p<W_1(x_0)/W(x_0)$;} \\
    W_1(x_0)+\int\limits_{\varepsilon_1}^{b}w_\phi(z+x_0|x_0)dz=p W(x_0), & \hbox{$p\geq  W_1(x_0)/W(x_0)$,}\label{l5}
  \end{array}
\right.
\end{equation}
find  $b$ corresponding to the  "transition channel"  $x_0 \to  b$.
\item  Draw a new number  $q\in(0,1)$  from a uniform distribution.
\item Reset time label  $t=t+\Delta t$  where  $\Delta t=-\ln q/W(x_0)$.
\item Reset $x_0$ to a new value $x_0+b$.
\item Return  to  the   step 2   and repeat the procedure  anew.
\end{enumerate}

\section{Statistics of random  paths for different jump-type processes}

In this section, we select the  jump-type  processes, which best demonstrate the capabilities of our algorithm of random paths generation.
Once suitable path ensemble data  are collected,    we   ca n reconstruct the pdf $\rho (x,t)$ dynamics   and next    verify whether
statistical  (ensemble)   features of generated random trajectories  are  compatible with the  asymptotic  a priori imposed upon solutions  of
the    master equation  (\ref{l1}).  That refers to  a control  of  an asymptotic behavior   $\rho (x,t) \to \rho_*(x)$ when  $t \to \infty$.

\subsection{Harmonic confinement}
Let us first consider an asymptotic invariant (target) pdf in the Gaussian form:
\begin{equation}
\rho_*(x)=\frac{1}{\sqrt{\pi}}e^{-x^2}.\label{l6}
\end{equation}
The corresponding $\mu $-family of transition rates reads
\begin{equation}
w_\phi(z+x|x)=C_\mu\frac{e^{-z^2/2+x z}}{|z|^{1+\mu}}.\label{l7}
\end{equation}
The C-codes for trajectory generating algorithm, \cite{available},  were employed to get  the  trajectory statistics
data for L\'{e}vy drivers with $\mu=0.5, 1, 1.5$.

\begin{figure}
\centerline{\includegraphics[width=0.5\columnwidth]{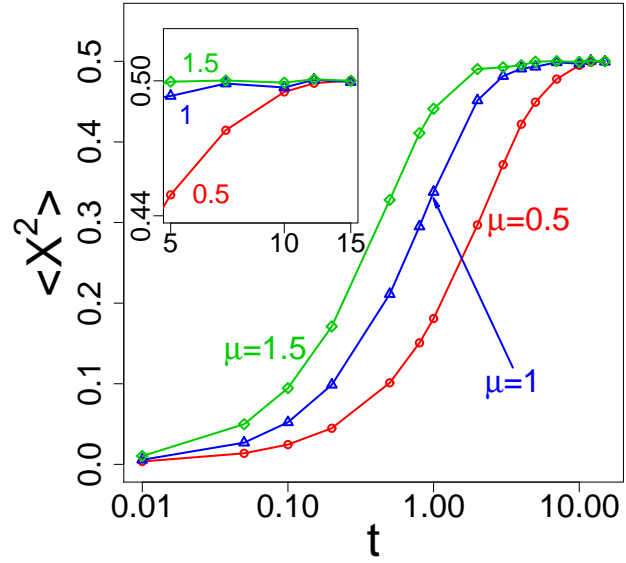}}
\caption{Gaussian target: Time evolution of  the pdf   $\rho (x,t)$  second moment   for  75 000 trajectories.
Inset visualize the oscillations smoothing in the asymptotic regime for $10\leq t\leq 15$; figures near curves correspond to $\mu$ values.}
\label{rys}
\end{figure}

\begin{figure*}
\centerline{
\includegraphics[width=0.3\columnwidth]{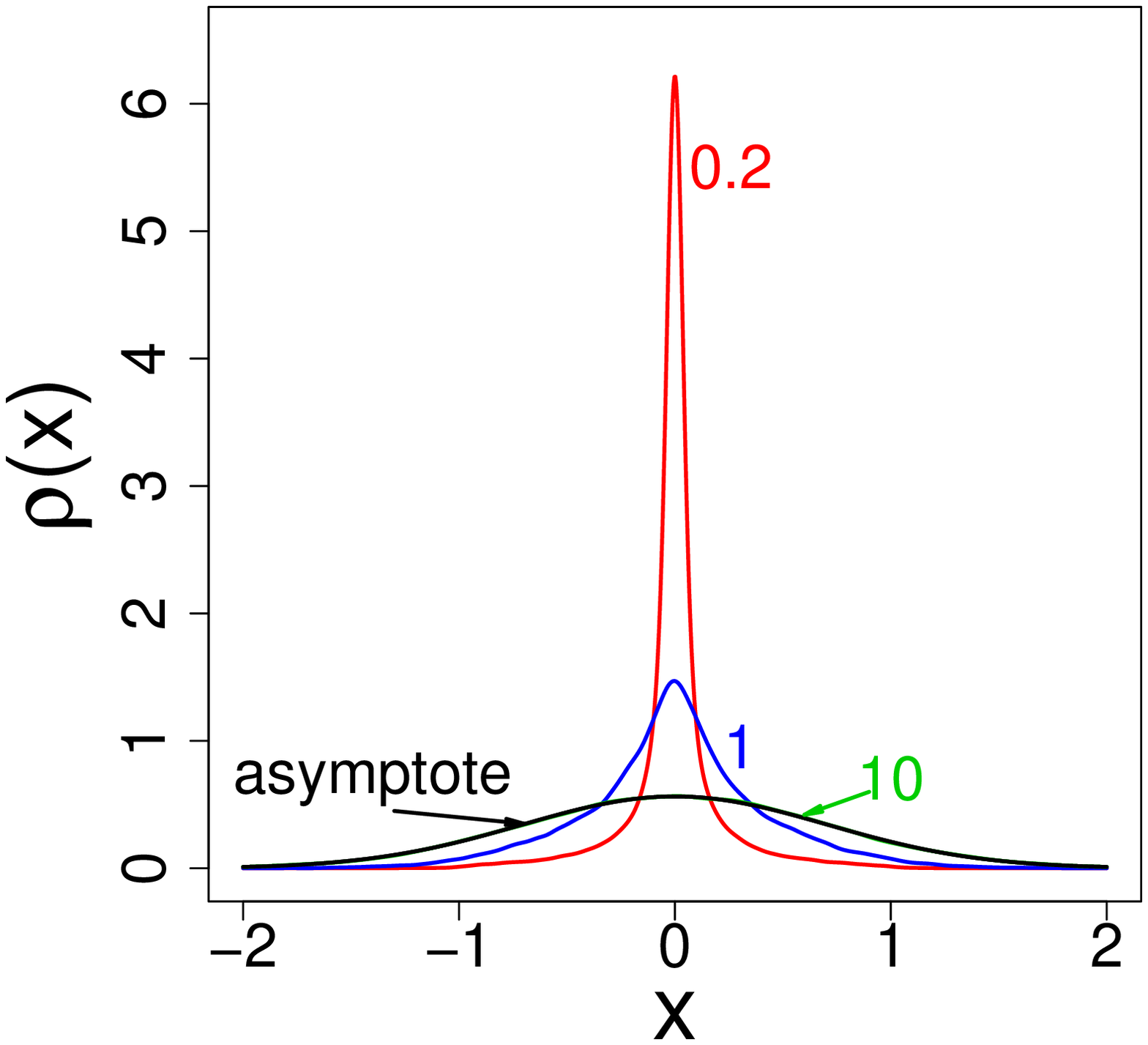}
\includegraphics[width=0.3\columnwidth]{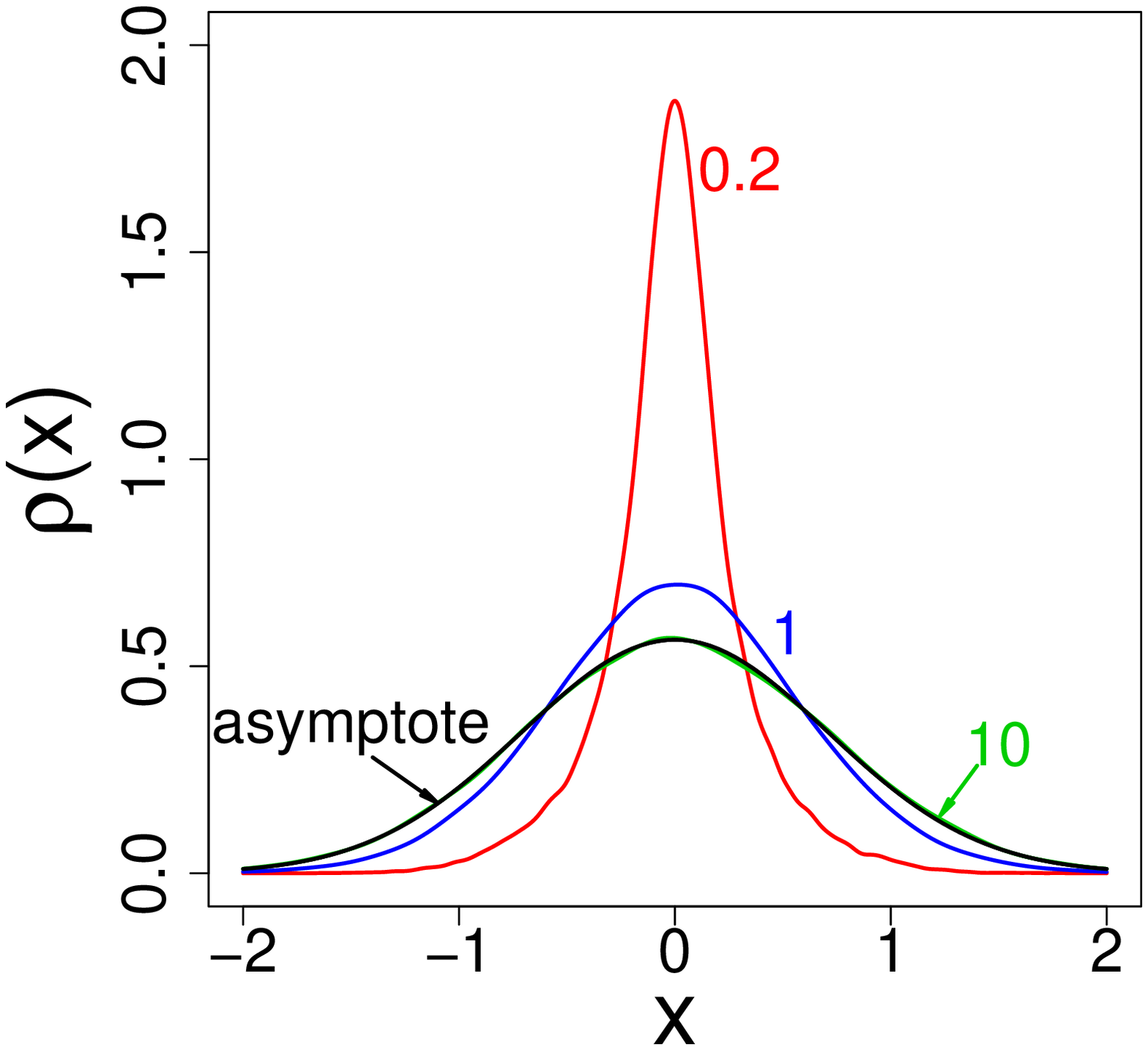}
\includegraphics[width=0.3\columnwidth]{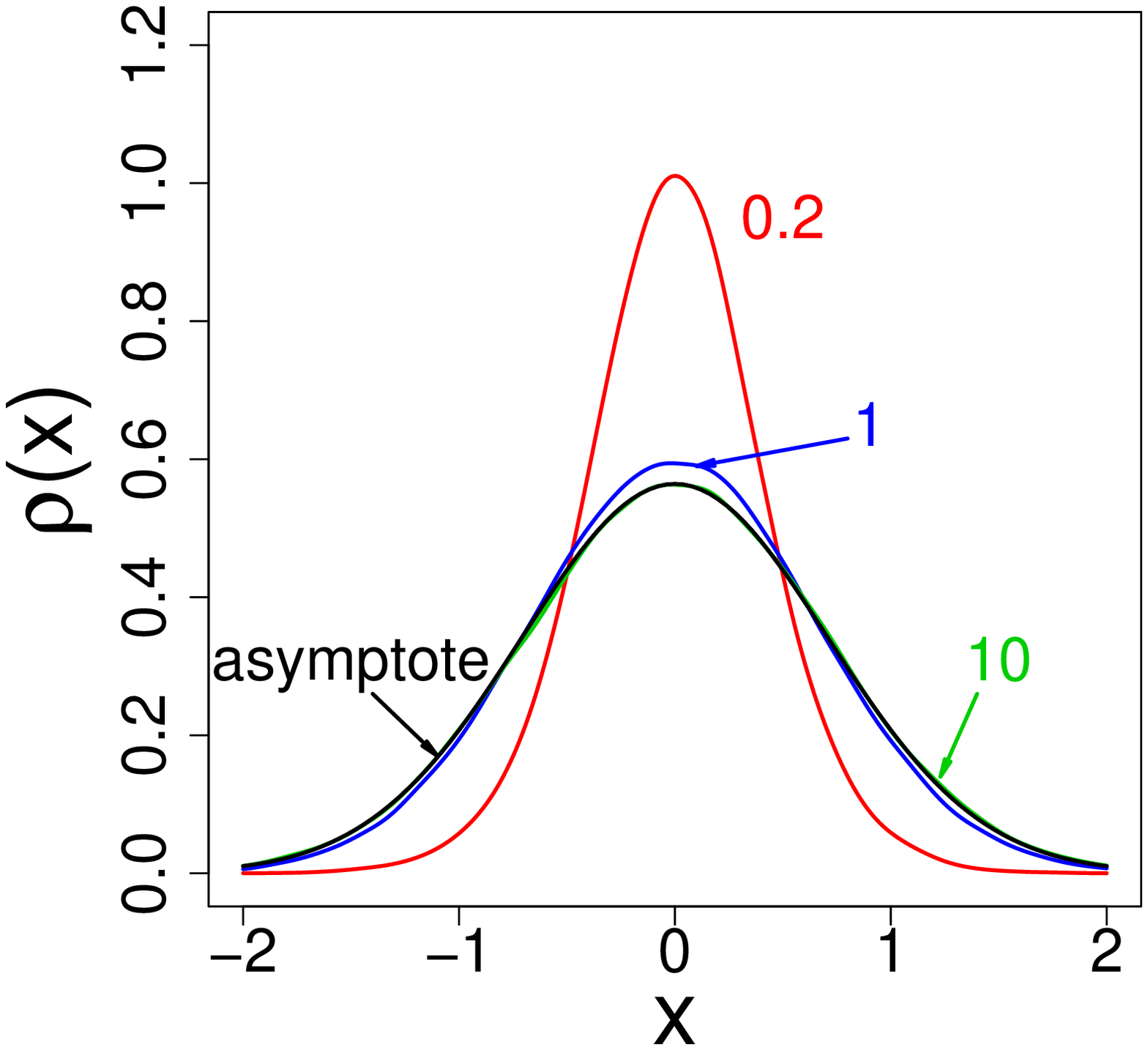}}
\caption{Gaussian target: Time  evolution of $\rho (x,t)$  inferred from  75 000 trajectories:  $\mu=0.5$ (left panel), $\mu=1$ (middle panel) and $\mu=1.5$ (right panel).
All trajectories originate from $x=0$,  i.e. refer to the $\delta(x)$-type initial probability distribution.}
\label{rys2}
\end{figure*}

The results of our numerical simulations are reported  in Figs. 1 and 2. We note, that in Fig. 1 the second moment oscillates near its equilibrium value $1/2$.  A numerical convergence to $<X^2>=$$1/2$ is consistent  with an analytic equilibrium value of the second moment  of  the chosen $\rho_*(x)$ \ref{l6}. The rate of this convergence is higher for larger $\mu \in (0,2)$. Clearly, for small $\mu$ the big jumps are frequent which  enlarges  the inferred time intervals  $\Delta t$ in the Gillespie's algorithm. Thus,  the relaxation  to equilibrium is slow. It gets   faster for  larger $\mu $, when big jumps are rare and time intervals  $\Delta t$  are  generically very small.

Fig. 2 displays a probability density evolution, inferred from the ensemble  statistics of   $75 000$  trajectories.
 All of them have started form the same point $x=0$.  Although the data  fidelity grows with  the number of contributing paths,
 we have not found significant  qualitative differences to justify a presentation of  data for 100 000, 200 000, 250 000 and more trajectories.
  The relaxation  time rate dependence on  $\mu $ is clearly visible as well. It suffices to analyze  differences between three curves for $t=0.2$ and/or  $t=1$. We observe a conspicuous lowering of their maxima  with  the growth
  of $\mu $ (take care of different scales on the vertical axes on Fig.2 panels).
The simulated  pdfs at $t=10$  are  practically indistinguishable from an exact analytical  asymptotic pdf \ref{l6}.
The  convergence of $\rho (x,t)$   towards    $\rho_*(x)$  appears  to be  relatively fast irrespective of the chosen $\mu $-driver.

Although our reasoning is definitely path-wise and all data have been extracted from simulated   trajectory ensembles,  our focus was on  the inferred dynamics of $\rho(x,t)$.
We  do not reproduce  figures  visualizing   generic sample paths.  However, we indicate their (paths)  distinctive  qualitative typicalities,  if    one would  compare  e.g.
   motion scenarios corresponding to stability indices  $\mu  = 1/2$, $\mu  =1$ and $\mu = 3/2$,    The structural impact    (probability of occurrence)  of larger against smaller
   jumps, even on the  visual level,   conforms  with  standard simulations of L\'{e}vy stable sample paths (with no forces or potentials involved), c.f. \cite{janicki}.

\subsection{Logarithmic confinement}
\subsubsection{Quadratic Cauchy target}

Let us consider  a long-tailed asymptotic pdf which is a special  $\alpha =2$ case of the  one-parameter
$\alpha $-family of equilibrium (Boltzmann-type) states,  associated with a logarithmic potential  $\Phi (x) \equiv  \alpha \ln (1+x^2)$, $\alpha  > 1/2$, see \cite{gar,gar0,gar1,gar2} :
\begin{equation}
\rho_*(x)=\frac{2}{\pi}\frac{1}{(1+x^2)^2}.\label{l8}
\end{equation}
The transition rate \ref{l2} $w_\phi(z+x|x)$ for any $\mu \in (0,2)$  takes the form
\begin{equation}
w_\phi(z+x|x)=\frac{C_\mu}{|z|^{1+\mu}}\frac{1+x^2}{1+(z+x)^2}.\label{l9}
\end{equation}

\begin{figure}
\centerline{\includegraphics[width=0.5\columnwidth]{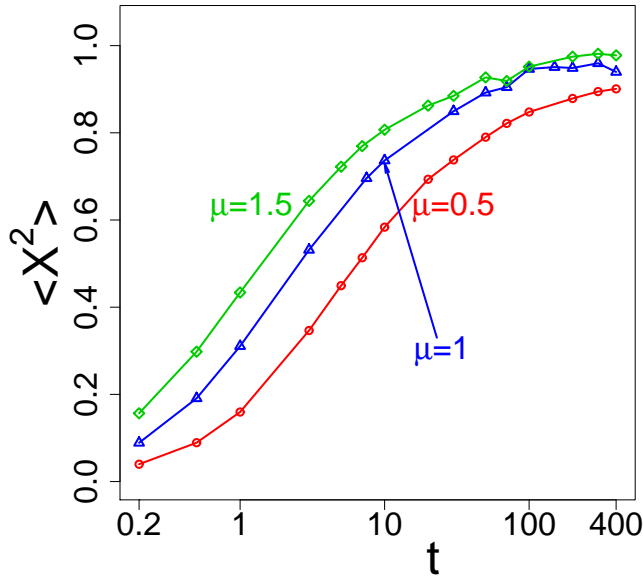}}
\caption{Quadratic Cauchy target: Time evolution of the pdf $\rho (x,t)$ second moment for 200 000 trajectories.}
\end{figure}

\begin{figure*}
\centerline{
\includegraphics[width=0.3\columnwidth]{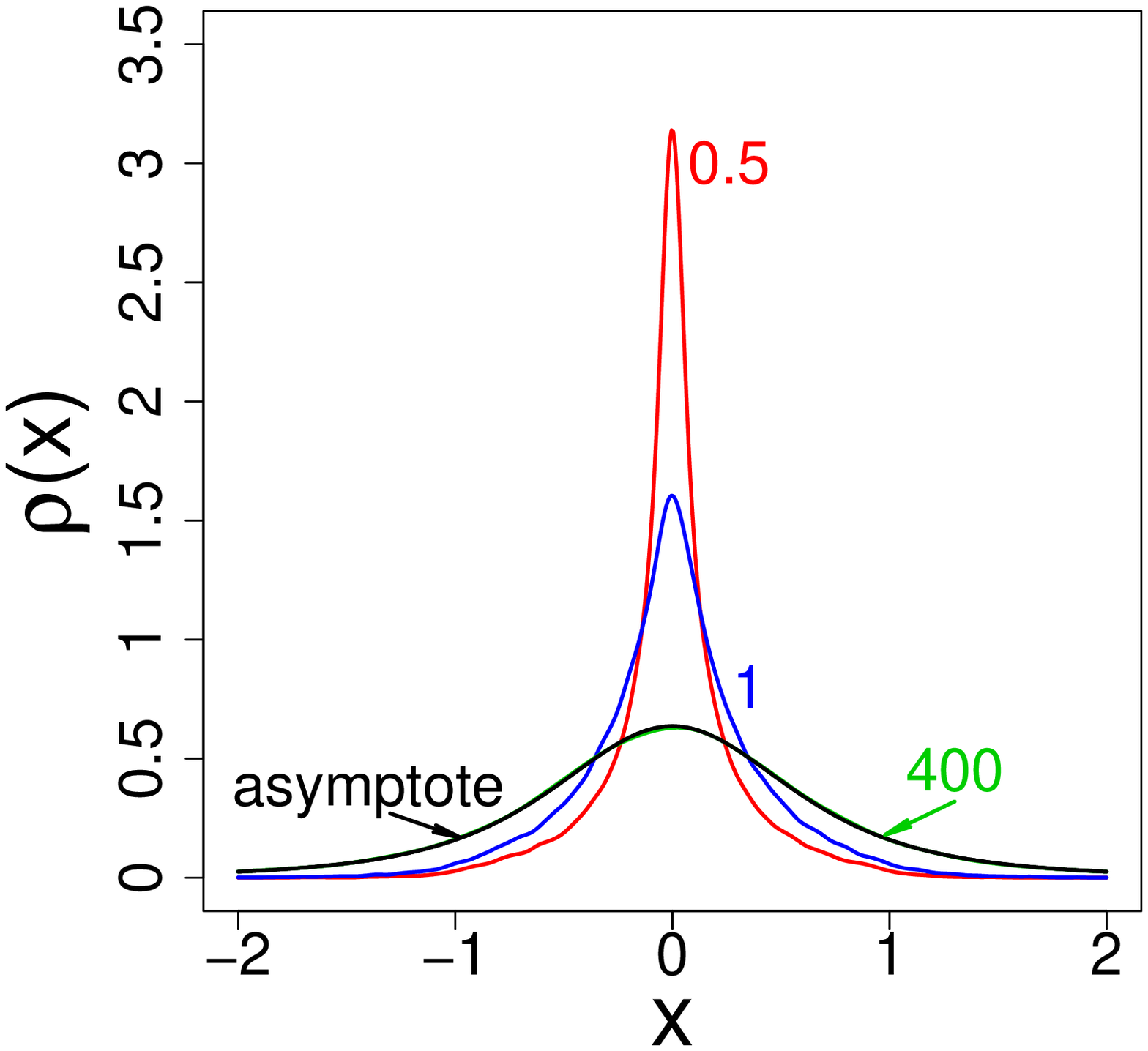}
\includegraphics[width=0.3\columnwidth]{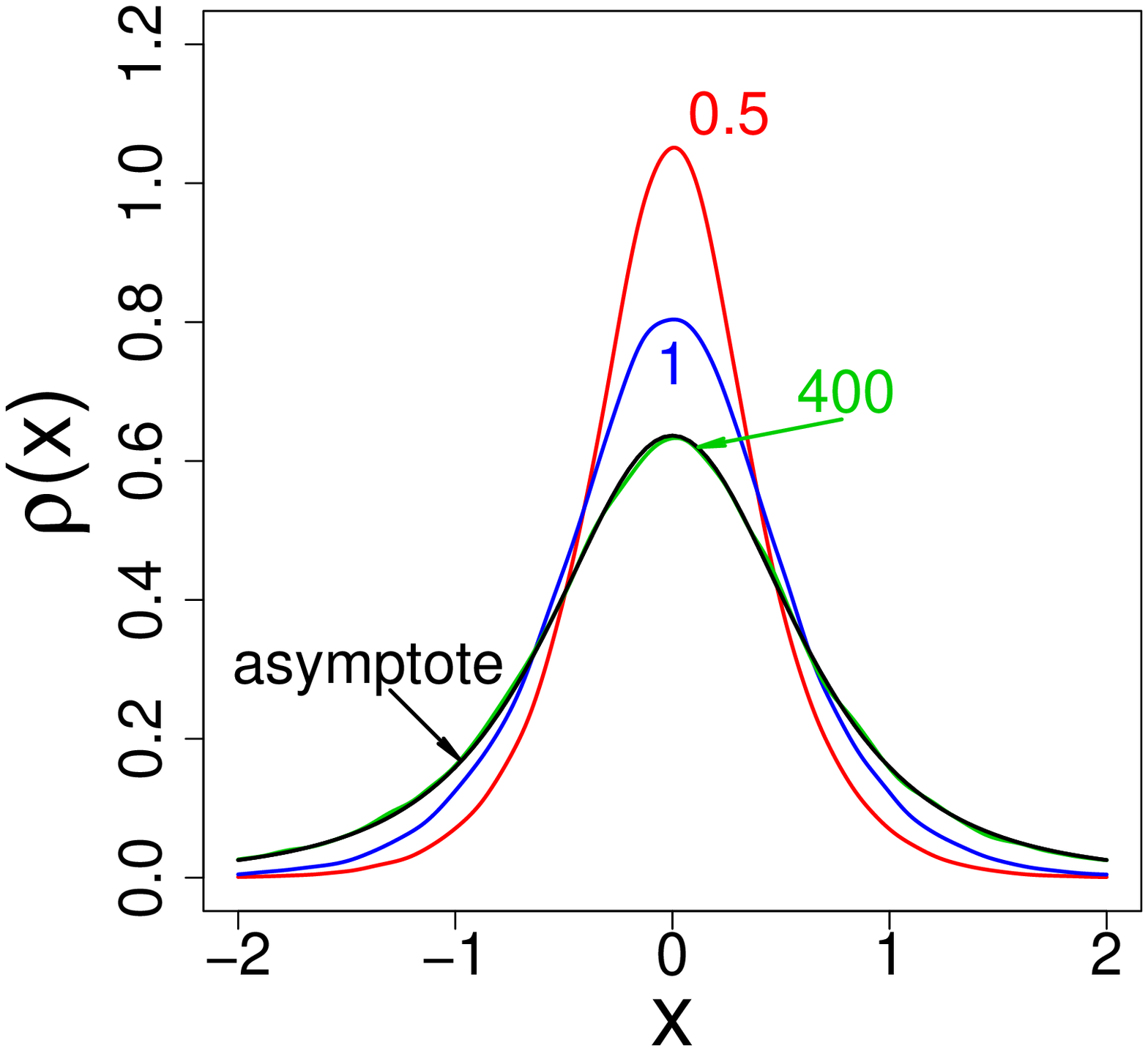}
\includegraphics[width=0.3\columnwidth]{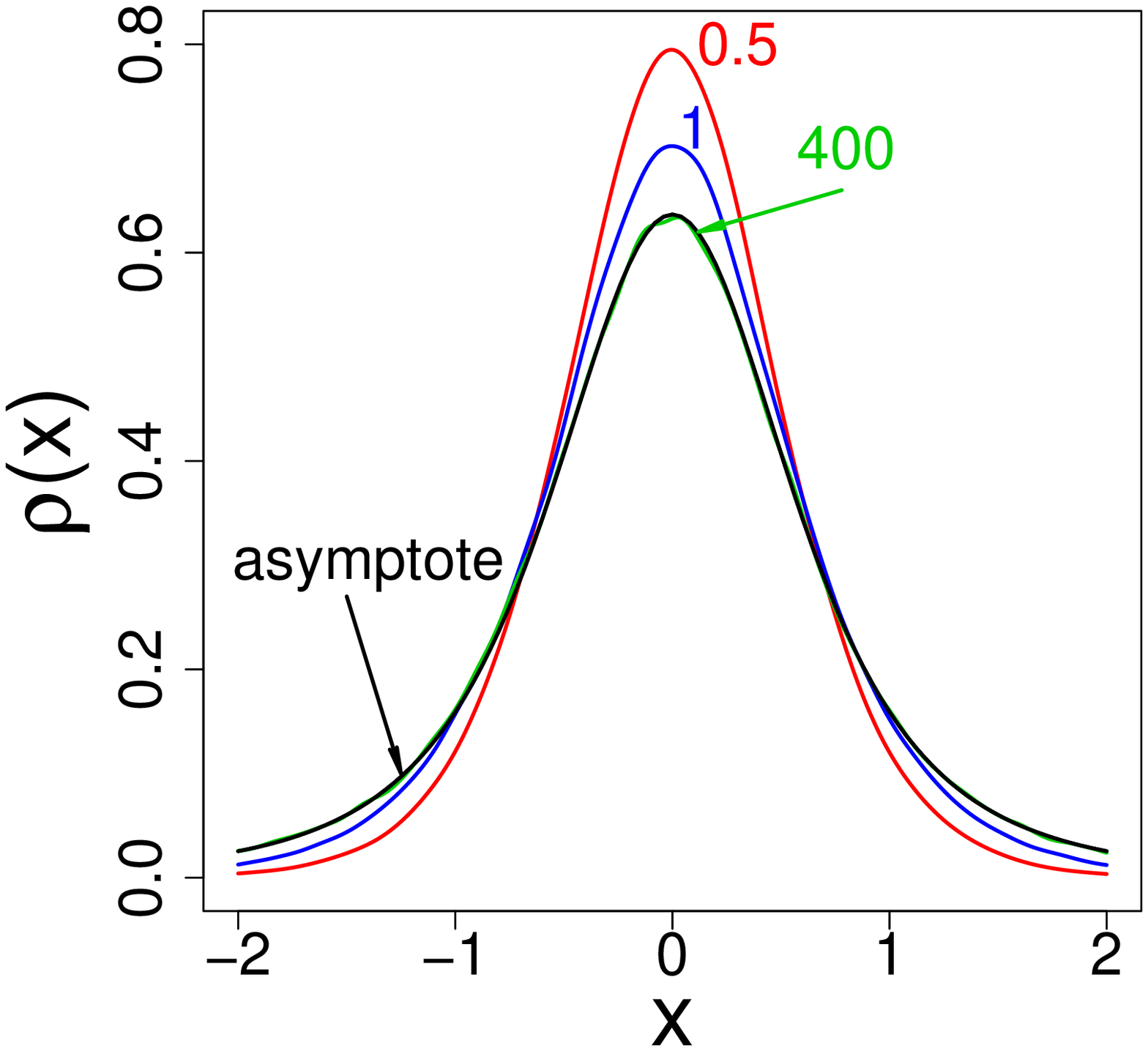}}
\caption{Quadratic Cauchy target: Time evolution of  $\rho (x,t)$ inferred from  $200 000$ trajectories for
$\mu=0.5$ (left panel), $\mu=1$ (middle panel) and $\mu=1.5$ (right panel).  All trajectories are started from $x=0$.
Note scale differences on vertical axes.}
\label{rys5}
\end{figure*}

Simulation results for this case are displayed in Figs. 3 and  4. If we compare Fig. 3 with Fig. 1 we see the existence of small oscillations in the asymptotic regime about the value $1/2$.
Those  from Fig.1 were relatively small, while those on Fig. 3 are more noticeable. This is related to much slower decay of transition rates (\ref{l9})
  (determined by slow-decaying asymptotic pdf  (\ref{l8})),    as compared to those for Gaussian case  (\ref{l7}).

The second moment of the present  $\rho_*(x)$, (\ref{l8}), equals 1 and the convergence towards this value  is clearly seen in Fig. 3.
 This convergence is much slower than in the Gaussian  (harmonic confinement) case  which is not a surprise:  (\ref{l7}) and  (\ref{l9})
   indicate that the present  rate of convergence should be logarithmically slower. Fig. 5,  quite alike Fig. 2, convincingly
  demonstrates a convergence  of $\rho (x,t)$ to the asymptotic $\rho _*(x)$.
  For definitely large  times around $t=400$,   $\rho (x,t)$  and $\rho _*(x)$  become   practically indistinguishable.
  Similarly  to the Gaussian case, the rate of convergence becomes larger  with the growth of   $\mu \in (0,2)$.

\begin{figure}
\centerline{
\includegraphics[width=0.5\columnwidth]{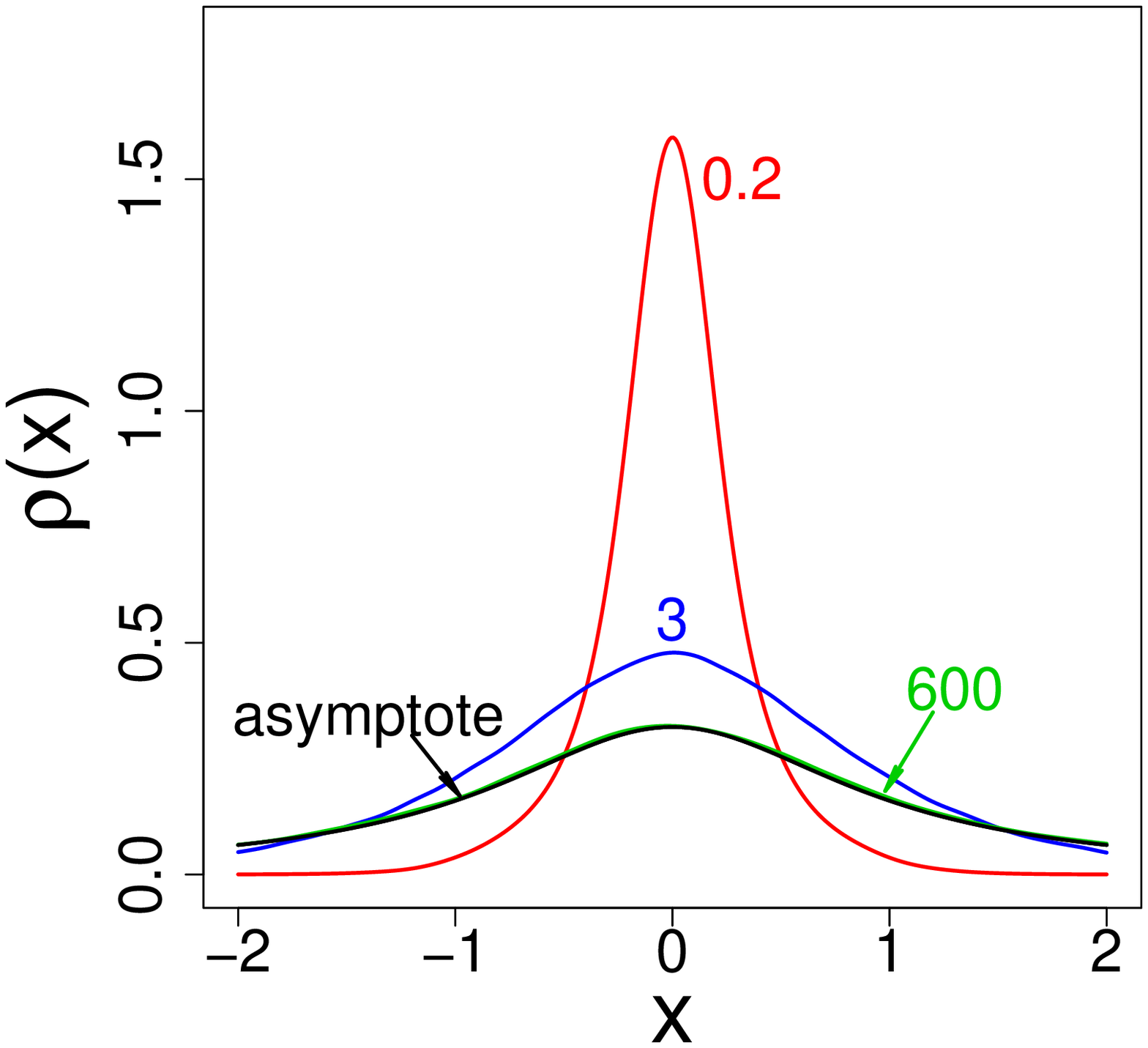}}
\caption{Cauchy target: Time evolution of pdf $\rho (x,t)$ for $\mu =1$ and $200 000$ trajectories, starting from $x=0$.}
\label{rys6}
\end{figure}

\subsubsection{ Cauchy target}

Now we consider  an asymptotic pdf  of the form :
\begin{equation}
\rho_*(x)=\frac{1}{\pi}\frac{1}{1+x^2}.\label{l20}
\end{equation}
In this case, the transition rate from $x$  to  $x+z$  reads
\begin{equation}
w_\phi(z+x|x)=\frac{C_\mu}{|z|^{1+\mu}}\sqrt{\frac{1+x^2}{1+(z+x)^2}}.\label{l21}
\end{equation}
We consider  Cauchy driver corresponding to $\mu=1$. In Fig. 5, we report the time evolution of the statistically inferred  $\rho (x,t)$ for different time instants. An approach to the asymptotic pdf  (\ref{l20}) is clearly seen, together with a convergence of a half-width to its asymptotic value 1.  It can be shown that the same convergence pattern is valid for cumulative probability distribution which approaches the asymptotic function $F(x)=\frac{1}{2}+\frac{\arctan x}{\pi}$.

\subsection{Locally  periodic confinement}

To set firm grounds for future research it is instructive to study our model for more complicated forms of confining potentials. In view of their physical relevance,  it is appealing to address an issue of confining (trapping) environments with a periodic spatial structure. This problem may be relevant to the random motion of defects in doped semiconductors and to phenomena like hopping conductivity.
Here, we encounter a major difficulty  with  integrability of the Boltzmann-type weighting function $\exp(-\Phi )$ as for periodic $\Phi$ the corresponding integral is divergent.
Periodicity and integrability can here be reconciled by means of locally periodic potentials that take a definite confining form (harmonic or polynomial) for larger values of $x\in R$. Let us consider the following asymptotic pdf
\begin{equation}
\rho_*(x)=\left\{
            \begin{array}{ll}
              \frac{1}{C} e^{-\sin^2(2\pi x)}, & \hbox{$|x|\leq 2$;} \\
              \frac{1}{C} e^{-(x^2-4)}, & \hbox{$|x|>2$,}
            \end{array}
          \right.
\end{equation}
where $C=3.032818$  is a normalization constant.
The transition rate from  $x$ to  $x+z$  reads
\begin{equation}
w_\phi(z+x|x)=\frac{C_\mu}{|z|^{1+\mu}}\exp{\left[(\phi(x)-\phi(x+z))/2\right]},
\end{equation}
where the  potential  $\phi$  has the following locally confining form
\begin{equation}
\phi(x)=\left\{
          \begin{array}{ll}
            \sin^2(2\pi x), & \hbox{$|x|\leq 2$;} \\
            (x^2-4), & \hbox{$|x|>2$.}
          \end{array}
        \right.
\end{equation}

\begin{figure}[h]
\begin{center}
\includegraphics[width=0.5\columnwidth]{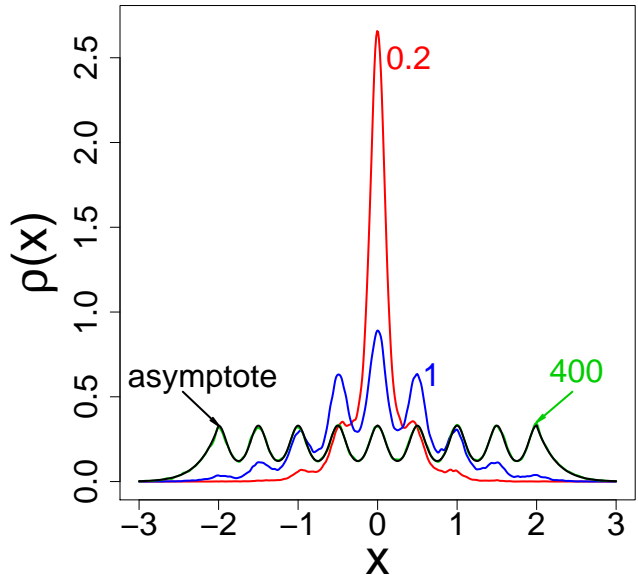}
\end{center}
\caption{Time evolution of $\rho (x,t)$ inferred from  $200 000$  trajectories at $\mu =1$. The  data for $100 000$ and $300 000$ trajectories
(not displayed) do not show qualitative differences.}
\end{figure}

Time evolution of the inferred  pdf $\rho (x,t)$ is reported in Fig. 6 for $\mu=1$. All sample trajectories were started form $x=0$ which  corresponds to the $\delta (x) $-type initial distribution. The probability density spreads out
with time in conformity with the trapping (confining) properties of the locally periodic enclosure
(environment or "potential landscape"). For large running times t=400 the trajectory statistics  produces data
 that are indistinguishable from those for the asymptotic pdf. We have checked that beginning  from about  100 000 trajectories, further accumulation of the trajectories number like e.g. 200 000 (displayed) and 300 000  (not displayed) for the data statistics is inessential. In such cases the curves are almost the same,  we  merely improve a fidelity of the statistics.

\section{Conclusions.}

If a random process does not admit the description in terms of a stochastic differential equation  (e.g. Langevin modeling), its direct numerical simulation
becomes  impossible  by means of  existing popular  algorithms. In the present paper, for the first time in the literature, we propose  a working method
to generate stochastic trajectories (sample paths) of a random jump-type process without resorting to any explicit (or numerical) solution of a stochastic
 differential equation.

  To this end we have  modified    the Gillespie algorithm \cite{gillespie,gillespie1},
normally devised for sample paths generation, if  transition rates occur  between a finite  number of states of  a system.
  The essence of our   modification is that we take into account the continuum of possible transition rates, thereby changing the finite  sums
  in the original Gillespie algorithm into  integrals.  The corresponding  procedure for  stochastic trajectories generation has  been
  changed accordingly.

  In other words, here we "extract" the background sample paths of a jump process, whose pdf    $\rho (x,t)$  is to obey  the generalized
  Fokker-Planck dynamics  \ref{l1},  \ref{l2}, . We emphasize once more  here, that  we have focused our attention on those background jump-type
  processes that  cannot be  modeled  by any stochastic differential equation of the  Langevin  type.  However, our ultimate  result was a reconstruction of the
  pdf dynamics  from the path-wise data.  Thus,  our modfication  of the  Gillespie's algorithm  serves two purposes:  (i)  getting access to  path-wise data
  and  (ii) reconstructing the pdf $\rho (x,t)$ dynamics, compatible with  the master equation in question,  without resorting to other  methods of   its solution

Although heavy-tailed  L\'{e}vy stable drivers were involved in the present considerations,
 we have clearly confirmed that an enormous variety of stationary  target  distributions is  dynamically accessible
   in each particular $\mu \in (0,2)$  case. That comprises not only a standard Gaussian pdf, discussed usually in relation
   to the Brownian  motion like Wiener process. We can reach Cauchy pdf in terms of any $\mu \neq 1$ driver as well,
    provided a steering environment is properly devised. In turn, the Cauchy driver  in a proper  environment
     may lead to an asymptotic pdf with a finite number of moments,  the Gaussian case included (\cite{gar}).

An example of the locally periodic environment has been considered as a toy model for more  realistic  physical systems.
We mention  possible  generalizations of our method to  the  Brownian motor concept  (see, e.g., Ref. \cite{browmot} for recent review)
 to include a  non-Gaussian jumping component. In those systems it  is   the  properly tailored periodic "potential landscape"
  which enforces a   conversion  of  a homogenous   stochastic process (Brownian motion for reference)
  into the directed motion of  particles at nanometer scales.
  It is closely related to the problem of so-called
 sorting in periodic potentials \cite{presort}. Other problems to be addresses  concern  ultracold atoms in optical lattices subject to
 random potentials \cite{coldopt}, which are promising not only from purely scientific point of view, but also for many
 technological applications.

 We note  that the  theoretical description of the  above   mentioned topics  relies  essentially on the
  Langevin-like equation  input.  Our approach  offers prospects for generalizations, where
 systems with non-Langevin response to external potentials may come  into  consideration, along with  more traditional ones.
 What we actually  need to implement our version of  Gillespie's algorithm   is the knowledge of
  jump  transition rates of those random  systems only.

Our preliminary work (in progress) shows that an extension of our algorithm to higher dimensions is operational. In particular, the planar case is worth further analysis, possibly with more realistic "potential landscapes".
The direct numerical modeling in three-dimensional case can also be  important, for instance, in the investigations of the dynamics of amorphous materials.
 Theoretical studies of their physical properties have a longer history, starting in the eighties and even earlier \cite{bin86,hoh}.
 These materials may comprise conventional glasses (see, e.g., \cite{par}), so-called spin  \cite{bin86,par} and orientational (for example, dipole) glasses \cite{hoh} as well as virtually any disordered solid like doped semiconductors.
  In these substances, the random motion of defects can be regarded as real jump-type process to be modeled numerically. Moreover, if defects have only rotational degrees of freedom (like spins or electric/elastic dipoles),
  their dynamics can be well represented by some effective jump-type process. Such representation can be complementary to existing Monte-Carlo modeling techniques which for disordered systems are very computer intensive.
  For instance, we hope to tackle the issue of interplay between exponential and long-time (logarithmic or stretched-exponential) relaxation in these substances.
  Also, the 3D modification of our algorithm can be used in the simulation of switching processes in disordered ferroelectrics \cite{my11}. Similar fictitious jump-type process can be assigned to the dynamics of   inhomogeneous
   broadening of resonant lines \cite{sto}. This
    broadening occurs in condensed matter and/or biological species, in a number of spectroscopic manifestations like the electron paramagnetic resonance, nuclear magnetic resonance,
    optical and neutron scattering methods. It  arises due to random electric and magnetic fields, strains and other perturbations from defects in a substance containing the centers whose resonant transitions
     between energy levels are studied. Here also, the standard approaches cannot describe all the details of experiment and we hope that our algorithm can improve the situation.

Let us finally note that in the 3D problems admitting full spherical symmetry (like spin glasses with random exchange interactions), the problem becomes reducible to 1D case so that our algorithm might work in its present form.

\textbf{Acknowledgement}:  P. G. would like to thank  Professor  I.  M. Sokolov for  focusing our attention on the Gillespie's algorithm.  All numerical  simulations were completed by means of the facilities of
the  Platon - Science Services Platform of the Polish Pionier Network.

\end{document}